# Multilayer C$_2$N: Effect of Stacking Order and Number of Layers on Bandgap and Its Controlled Electronic Properties by External Electric Field

*Ruiqi Zhang[a], Bin Li[a] and Jinlong Yang[a, b, *]*

[a] Hefei National Laboratory for Physical Sciences at Microscale, University of Science and Technology of China, Hefei, Anhui 230026, China

[b] Synergetic Innovation Center of Quantum Information & Quantum Physics, University of Science and Technology of China, Hefei, Anhui 230026, China

**ABSTRACT**

Successful synthesis of the nitrogenated holey two-dimensional structures C$_2$N (*Nat. Commun.* **2015**, *6*, 1–7) using simply wet-chemical reaction offer a cost-effective way to generate other 2D materials with novel optical and electronic properties. Using the few-layer C$_2$N as models, we have performed an ab initio study of electronic properties of layered C$_2$N. Band gaps of this system exhibit monotone decreasing as the number of layers increase. And a direct-gap to indirect-gap transition at the bulk C$_2$N. Besides, when we apply an out-of-plane electric field on few-layer C$_2$N, the band gap of multilayer C$_2$N will be decreased as the electric field increased and a semiconductor-semimetal transition will happen for five-layer $C_2N$ under an appropriate electric field , whereas the band gap of monolayer $C_2N$ is unchanged under electric field. Owing to their tunable bandgaps in a wide range, layers $C_2N$ will have tremendous opportunities to be applied in nanoscale electronic and optoelectronic devices.

1. INTRODUCTION

In recent years, two-dimensional (2D) materials are considered emerging ingredients for the future of nanoelectronics and optoelectronics applications due to their fascinating electronic, mechanical, optical or thermal properties.[1–3] Graphene, a hexagonal lattice of carbon atoms, have been attracted intensive research since its isolation one of the promising candidates for future applications in nanoscale electronics.[4–6] However for graphene, the massless Dirac-fermion behavior make it possess extremely high mobility, but the absence of a fundamental band gap severely limits its applications in field-effect transistors.[5,7,8] Then extensive efforts have been devoted to solve the problem of opening a gap in different graphene nanostructures.[9] At the same time, a new class of 2D materials has been studied, such as few-layer transition metal dichalcogenides (TMDs),[10–12] and in particular $MoS_2$ does possess a direct bandgap of ~1.8 eV.[10] Although monolayer $MoS_2$ has recently been used to fabricate a FET, the low carrier mobility[13] of ~200 $cm^2$ $V^{-1}$ $s^{-1}$ limit its wide application in electronics. Recently, monolayer black phosphorous (phosphorene), with high mobility, high in-plane anisotropy and direct-bandgap semiconductor, was isolated from the bulk black phosphorus,[14–17] which has immediately received considerable research attention.[18–22] Although $MoS_2$ and phosphorene simultaneously satisfy appreciable band gap and high carrier mobility, their puckered lattices cannot strictly confine the movement of the carrier within the 2D surface.[23] Thus, a search for 2D planer materials with a suitable stable bandgap is still ongoing.

Recently, a thinnest layered 2D crystal named $C_2N$, with uniform holes and nitrogen atoms, can be simply synthesized via a bottom-up wet-chemical reaction. Furthermore, a FET based on layer $C_2N$ with an high on/off ratio of $10^7$ was fabricated and $C_2N$ possess a optical bandgap of ~1.96 eV.[24] These results may make layer $C_2N$ a very promising candidate material for future applications in nanoscale electronics and optoelectronics.

Here, we perform first-principles to investigate the electronic structures of few-layer and bulk $C_2N$ resulting in several important findings. First, we found the monolayer $C_2N$ with a direct band gap of ~1.66 eV at the Γ points based on our calculations. Besides, we explored the electronic properties of few-layer $C_2N$ as a function of number layers. Our result show taht layer-$C_2N$ exhibits a sizable and novel direct band gap, which is decreased as the number layers increased. What is more, when we explored how multilayers $C_2N$ responses to a vertical electric field, we found the band gap of layer-$C_2N$ are determined by the number of layer and the electric field intensity. With the number of layer and the electric field intensity increased, few-layer $C_2N$ can be transform from a semiconductor to semimetal. All in all, the band gap of few-layer $C_2N$ can be tuned in a relatively wide range, increasing the tunability for their potential application in nanoelectronics.

2. **METHODS**

In this study, our first-principles calculations are based on the density functional theory (DFT) implemented in the VASP package.[25,26] The generalized gradient approximation of Perdew, Burke, and Ernzerhof (GGA-PBE)[27] and projector augmented wave (PAW) potentials are used. In all computations, the kinetic energy cutoff are set to be 520 eV in the plane-wave expansion.

For the geometry optimization, 5×5×1 and 5×5×8 Monkhorst-Pack k-meshes are adopted for the bulk and few-layer $C_2N$, respectively. A large value ~15 Å of the vacuum region is used to avoid interaction between two adjacent periodic images. All the geometry structures are fully relaxed until energy and forces are converged to $10^{-5}$ eV and 0.01 eV/Å, respectively. Dipole correction is employed to cancel errors of electrostatic potential, atomic forces and total energy, caused by periodic boundary condition. Effect of van der Waals (vdW) interaction is accounted for by using empirical correction method proposed by Grimme (DFT-D2),[28] which is a good description of long-range vdW interactions.[22,29,30] As a benchmark, DFT-D2 calculations give an interlayer distance of 3.25 Å and a binding energy of -25 meV per carbon atom for bilayer graphene, consistent with previous experimental measurements and theoretical studies.[31,32]

**RESULTS AND DISCUSSION**

To have a thorough knowledge of 2D $C_2N$, we first studied the geometric properties of monolayer $C_2N$. The atomistic ball-stick models of monolayer $C_2N$ with a 2×2 supercell are illustrated in Fig. 1(a) and 1(b). Their plane structures are fully relaxed according to the force and stress calculated by DFT within the PBE functional, while the few-layer $C_2N$ is obtained by the DFT-D2 method.

The equivalent lattice parameters of monolayer $C_2N$ is 8.277 Å, generating the in-plane covalent bond lengths as C-C of 1.430 Å and C-N of 1.336 Å. Carefully looking at the geometric structures, we found there are 12 C atoms and 6 N atoms in a unite cell and uniform holes in the layer $C_2N$. By performing DFT-GGA calculations, it shown that a single layer $C_2N$ is a semiconductor with a direct band gap of 1.66 eV at the Γ points, which is depicted in Fig. 1(c). At the same time, we plot the isosurfaces of band decomposed charge density corresponding to the valence band maximum (VBM) and conduction band minimum (CBM), which is shown in Fig. 1(c) and (d), respectively. It is clear that the distribution of VBM and CBM is derived from different atoms: the former mainly originates from the nitrogen $P_z$ states and the latter is mainly from the C=C antibonding π states.

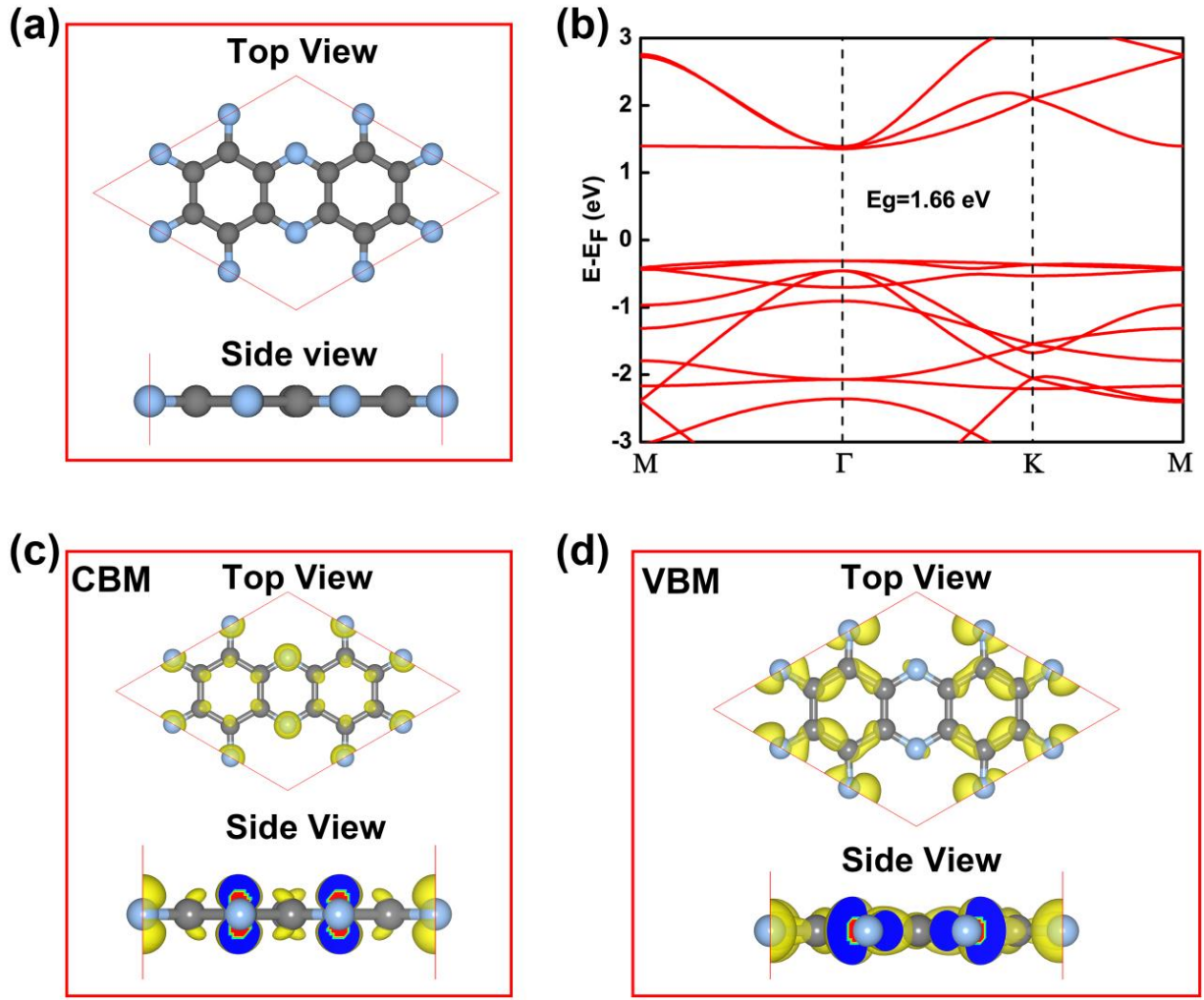

Figure 1. (a) Top view and side view of the atomic structure of monolayer C2N. A 2 × 2 supercell is taken for brevity; (b) Band structures of monolayer C2N calculated by PBE; (c,d) are the charge density corresponding to the CBM and VBM for monolayer, respectively. The isovalue is 0.003 e/Bohr$^3$. The grey and silvery ball represent C atoms and N atoms, respectively.

Γ (0.0, 0.0, 0.0), M(0.5, 0.5, 0.0), and K (1/3, 1/3, 0.0) refer to special points in the first Brillouin zone.

To explore the bilayer $C_2N$, we considered three kinds of high symmetry stacking structures, namely, AA-, AB-, and AC-stacking. As shown in Fig. 2(a), for the AA-stacking, the top layer is directly stacked on the bottom layer. As for the AB-stacking, it is the C rings of the bottom layer that are under the center of the top layer holes. However, for the AC-stacking, it can be viewed as shifting the top layer of the AA-stacking by half of the cell along the basis vector, which result in the hexatomic rings composed by C and N atoms are just right under the center of the top layer holes. Our total energy calculations based on the PBE functional indicate that AB-stacking is the most favorable, which is 16 and 3 meV/atom lower than that of AA- and AC- stacking, respectively. Besides, we also note that the computed lattice constants differ slightly.

The PBE electronic band structures of AA-, AB-, and AC-stacked bilayer $C_2N$ are shown in Fig. 3(d)−(f). Clearly, the direct-gap feature is retained regardless of the stacking orders. Both the VBM and CBM are located at the Γ point. Among the three bilayers with different stacking orders, the AB-stacked has the widest band gap of 1.49 eV. While the band gap of AA-, AB-stacked bilayer $C_2N$ is 1.34 eV and 1.21 eV, respectively. Hence, it is different stacking-order results in different interactions between layers that makes different interaction strength and bandgap.

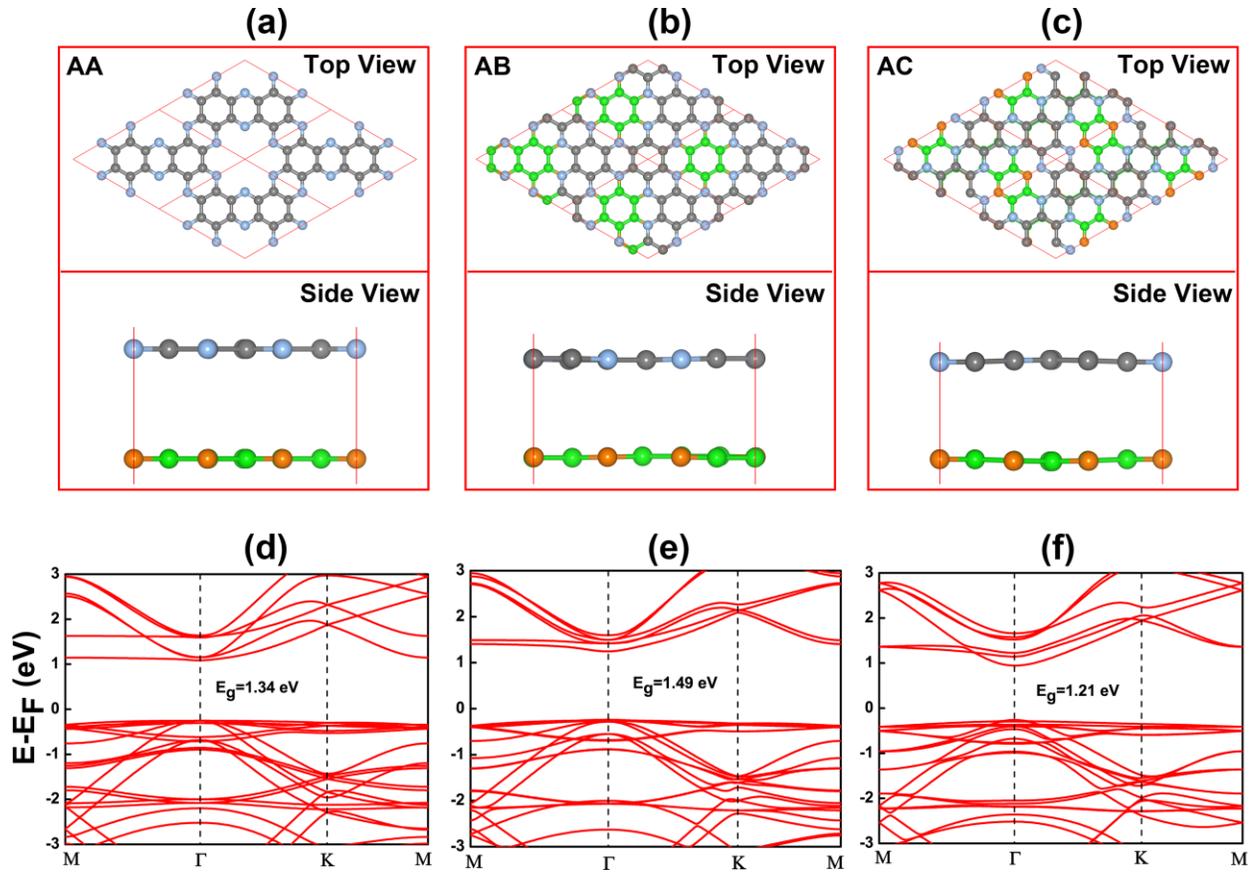

Figure 2. Three stacking structures of bilayer $C_2N$. (a,b,c) Top view and side view of AA-, AB-, and AC-stacking. A 2 × 2 supercell for top view is taken for brevity; (d,e,f) the electronic band structures of AA-, AB-, and AC-stacked bilayer $C_2N$, respectively. C atoms in different layers are represented by grey and green balls, respectively. And N atoms in different layers are represented by blue and orange balls, respectively. Γ (0.0, 0.0, 0.0), M(0.5, 0.5, 0.0), and K (1/3, 1/3, 0.0) refer to special points in the first Brillouin zone.

Then, we considered tow stacking ways for trilayer $C_2N$ based on AB- stacking: one is ABA stacking and the other one is ABC stacking. The difference between them is C can be considered as shifting A by half of the cell along the basis vector. The atomic structure of trilayer $C_2N$ are shown in Fig.4 (a) and (d), respectively. According to our total energy calculations, it is the ABC-stacking that has the more stable structure, which is 2 meV/atom lower than that of ABA-stacking. Then, their band structures are depicted in Fig. 4(b) and (e), respectively. Obviously, their band structures are almost same and suggest trilayer $C_2N$ is a semiconductor with a direct bandgap. All their CBM and VBM are located in Γ point and the isosurfaces of the charge density corresponding to them are shown Fig. 4 (c) and (d).

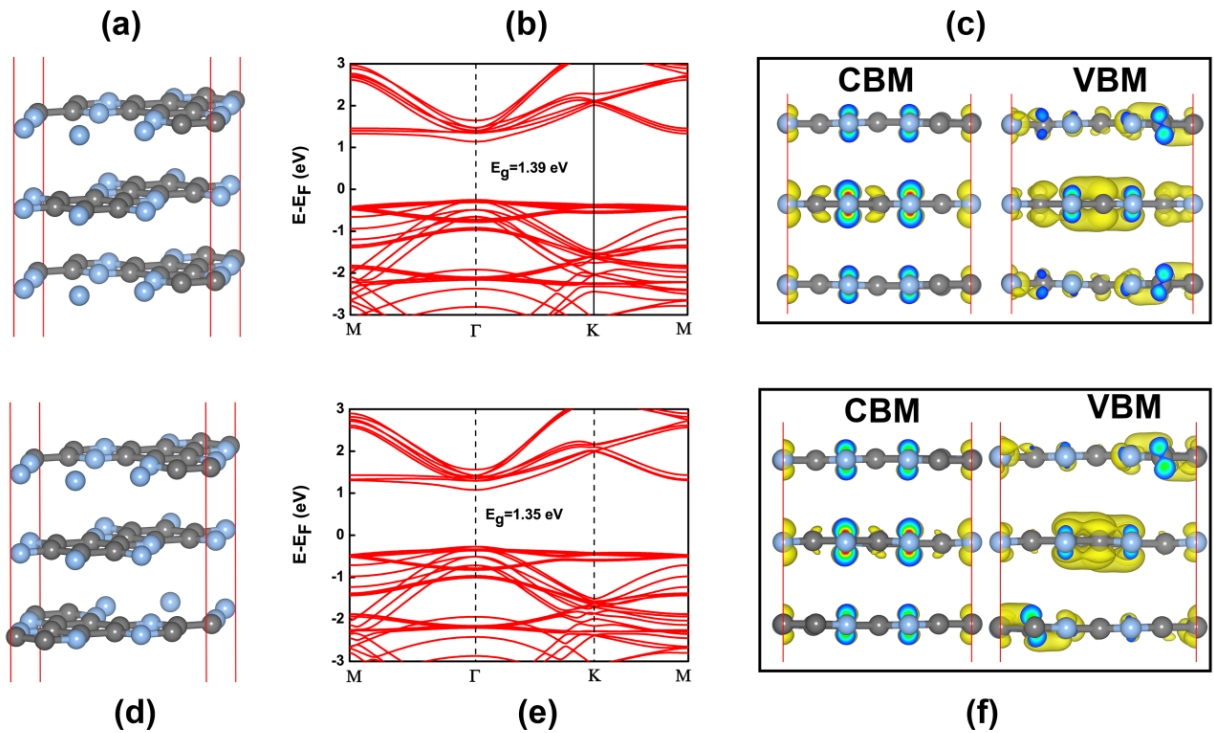

Figure 3. (a,d) The atomic structure of ABA- and ABC-stacked trilayer $C_2N$, respectively; (b,e) the electronic band structures of ABA- and ABC-stacked trilayer $C_2N$; (c) Side view of the charge density corresponding to the CBM and VBM for ABC-stacked. The isovalue is 0.003 e/Bohr$^3$. The grey and silvery ball represent C atoms and N atoms, respectively. Γ (0.0, 0.0, 0.0), M(0.5, 0.5, 0.0), and K (1/3, 1/3, 0.0) refer to special points in the first Brillouin zone.

Then, we build multilayer even to bulk C$_2$N according to ABC–stacking. The calculated lattice (a) constant and the interlayer distance (Δc) are list in Table.2. The lattice parameter a decrease by 0.022 Å on passing from monolayer to bulk C$_2$N, whereas the interlayer distance Δc decrease by 0.054 Å from bilayer to bulk phase, which suggest the thicker the few-layer C$_2$N system the stronger is the interlayer interaction. While, the length of chemical bonding of C-N and C-C almost keep the same in the different layers.

Table.1 Lattice Constants a and Δc (Interlayer Distance) and the in-Plane Covalent Bond Lengths of Few-Layer C2N Calculated Using DFT-D2

| N$_L$ | a (Å) | Δc (Å) | C-N (Å) | C-C(Å) |
|---|---|---|---|---|
| 1 | 8.328 |  | 1.336 | 1.429/1.470 |
| 2 | 8.321 | 3.185 | 1.336 | 1.429/1.468 |
| 3 | 8.315 | 3.156 | 1.335 | 1.429/1.466 |
| 4 | 8.315 | 3.157 | 1.335 | 1.429/1.466 |
| 5 | 8.313 | 3.150 | 1.336 | 1.429/1.466 |
| bulk | 8.306 | 3.131 | 1.335 | 1.428/1.465 |

Another effect of layer increasing reflects on the band structures. To explore the reflects, we examine the electronic properties of few-layer C$_2$N nanosheets and find that the band gaps decrease as the number of layers increase, which is shown in Fig. 5(c). The few-layer C$_2$N has a direct bandgap, tunable from 1.66 eV for a monolayer to 1.23 eV for a five-layer. However the bulk phase possess an in-direct bandgap with a value of 1.07 eV, whose atom atomic structure and band structure are shown in Fig.5 (a) and (b). Clearly, the CBM of bulk $C_2N$ is located in A point and VBM is in Γ point.

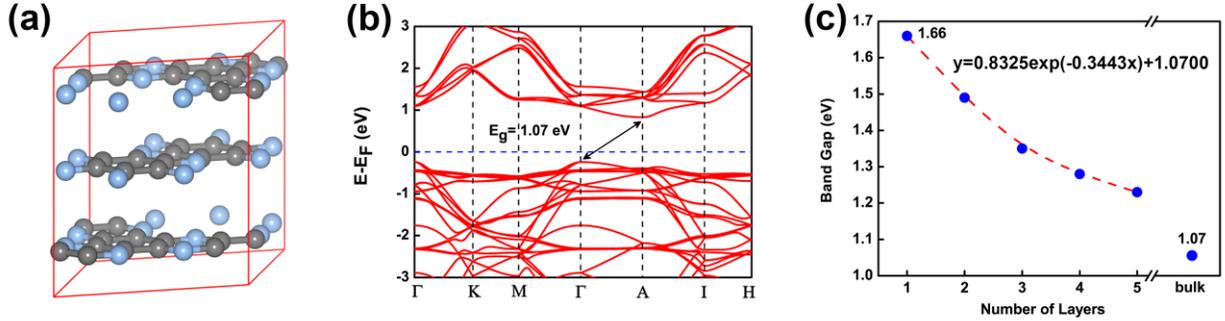

Figure 4. (a) Crystal structure of bulk $C_2N$; (b) Band structures of bulk $C_2N$ calculated by PBE.;(c) Evolution of the band gap of layer $C_2N$ as a function of layer number.

Previous theoretical and experiment studies had showed that the external electric field may affect the bandgap of 2D materials.[33–38] So it is interesting to study how few-layer $C_2N$ responses to a vertical electric field. Then we choose the monolayer to five-layer $C_2N$ as models to test. The electric field is applied perpendicular to the multilayer $C_2N$ surface, with a strength ranging from 0.1 to 0.5 V/Å. For the monolayer, its band gap is almost unchanged as the external electric field increases. In contrast, the band gap of few-layer $C_2N$ exhibit a monotonic decreasing relationship with increasing electrical field and close up as the electric field of 0.5 V/Å be applied in five-layer $C_2N$ based on our PBE calculation, which is illustrated in Fig. 5(a) and (b). It is noted that the direct bandgap structures of our model with increasing electrical field are all kept until five-layer $C_2N$ turn to a semimetal under the electric field of 0.5 V/Å. In order to explain why this transition happen in few-layer $C_2N$, we calculate their band structure. For bilayer $C_2N$, it is clearly shown in Fig. 5(c) that CBM is derived from first layer and VBM is come from second layer. However, for five-layer $C_2N$, CBM is derived from the first layer and VBM is come from the bottom layer, which is shown in Fig. 5 (b). Compared with Fig. 5(c) and Fig. 5(d), the bands of bilayer $C_2N$ are will kept in five-layer $C_2N$ under the same electrical. It is clearly that the CBM get closer to the VBM as the value of electric field increase. It can be understood as bands shift for different layers happened under electric field, which is shown in Fig. 5(e). Then, when we plot the isosurfaces of the charge density corresponding to CBM and VBM of five-layer $C_2N$ at Γ point. For example, when the strength of electric field get 0.4 V/Å, the CBM almost comes from the first layer and VBM from the bottom layer, which is consistent with their band structures.

The above-discussed band gap modulation of few-layer $C_2N$ under external electric field can be explained by Stark effect. Similar phenomenon has been observed in single-walled boron nitride nanotubes,[39,40] boron nitride nanoribbons[41,42] and $MoS_2$ bilayer.[43] When an external electric field is applied, the CBM and VBM in few-layer $C_2N$ become localized at the two different edges. Owing to the external electrostatic potential difference between the two edges, the bands belonging to different $C_2N$ layers are further split by the vertical electric field. As a result, the band gap is reduced with increased field strength and finally leads to a semiconductor to semimetal transition.

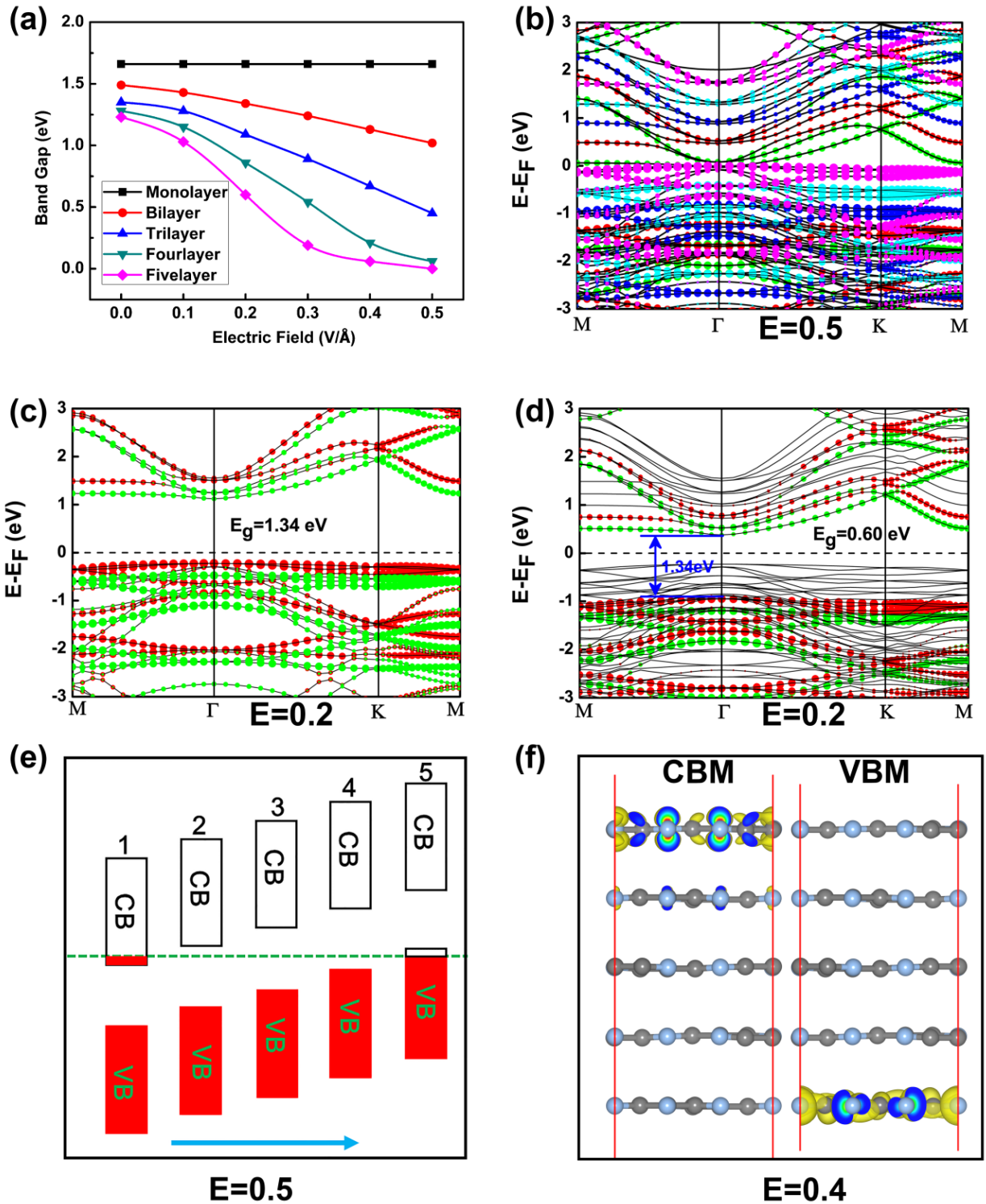

Figure 5. (a) Variations of band gap of layer C$_2$N under the external field; (b) Computed band structures (PBE) of five layer C$_2$N under electric field of 0.5 V/Å, bands of the first, second,

third, fourth and fifth layer are represented by green, red, blue, cray and magenta lines mark, respectively; Computed band structures (PBE) of (c) bilayer $C_2N$ under electric field of 0.2 V/Å, (d) five layer $C_2N$ under electric field of 0.2 V/Å, respectively. (e) A schematic view of bands shift for five-layer $C_2N$ under electric field of 0.5 V/Å, the green dotted line represents the Fermi level and the light blue arrow represents the direction of electric field. (f) the charge density corresponding to the CBM and VBM of five-layer $C_2N$ under the electric fields of 0.4 V/Å. The isovalue is 0.003 e/Bohr$^3$.

**IV. SUMMARY**

In conclusion, on the basis of DFT calculations, we have performed a theoretical research on the structural and electronic properties of few-layer $C_2N$. We have shown theoretically that few-layer $C_2N$ is a novel category of 2D direct band gap semiconductor, and the bandgap decreases as the number of layer increased. Owing to the band gap of multilayer $C_2N$ are determined by edge states, when an external electric field perpendicular to the few-layer $C_2N$ surface is applied, the bandgap of few-layer $C_2N$ also decreases with increasing the vertical electric field and can be tuned in a relatively wide range, while for monolayer the band gap is unchanged. For example, under the electric field of 0.5 V/Å, five-layer $C_2N$ will tune to a semimetal, which can be explained by Stark effect. Thus, our theoretical predictions suggest that layer $C_2N$ is very promising for optoelectronic applications, due to its tunable bandgaps by the number of layers and an external electric field, possible semiconductor to semimetal transition.

**AUTHOR INFORMATION**

**Corresponding Author**

\* E-mail: jlyang@ustc.edu.cn. Phone: +86-551-63606408. Fax: +86-551-63603748 (J. Y.).

**Author Contributions**

The manuscript was written through contributions of all authors. All authors have given approval to the final version of the manuscript.

**ACKNOWLEDGMENT**

This work is partially supported by the National Key Basic Research Program (2011CB921404), by NSFC (21121003, 91021004, 21233007, 21203099, 21273210), by CAS (XDB01020300), and by USTCSCC, SCCAS, Tianjin, and Shanghai Supercomputer Centers.